\documentclass[twocolumn,showpacs,preprintnumbers,amsmath,amssymb,floatfix,aps]{revtex4}
\usepackage{graphicx}

\begin{document}
\title{Phonons and Magnetic Excitations in Mott-Insulator LaTiO$_3$}
\author{M. N. Iliev$^1$, A.~P.~Litvinchuk$^1$, M.~V.~Abrashev$^2$,
V.~N.~Popov$^2$, J.~Cmaidalka$^1$, B.~Lorenz$^1$, R.~L.~Meng$^1$}
\affiliation{$^1$Texas Center for Superconductivity and Advanced
    Materials, and Department of Physics, University of Houston,
    Houston, Texas 77204-5002\\
$^2$Faculty of Physics, University of Sofia, 1164 Sofia, Bulgaria}
\date{July 18, 2003}
\begin{abstract}
The polarized Raman spectra of stoichiometric LaTiO$_3$ (T$_N =
150$~K) were measured between 6 and 300~K. In contrast to earlier
report on half-metallic LaTiO$_{3.02}$, neither strong background
scattering, nor Fano shape of the Raman lines was observed. The
high frequency phonon line at 655~cm$^{-1}$ exhibits anomalous
softening below T$_N$: a signature for structural rearrangement.
The assignment of the Raman lines was done by comparison to the
calculations of lattice dynamics and the nature of structural
changes upon magnetic ordering are discussed. The broad Raman
band, which appears in the antiferromagnetic phase, is assigned to
two-magnon scattering. The estimated superexchange constant
$J~=~15.4\pm0.5$~meV is in excellent agreement with the result of
neutron scattering studies.
\end{abstract}
 \maketitle
There is still debate on the role of orbital degrees of freedom in
the antiferromagnetism of LaTiO$_3$ and whether orbital ordering
exists in the antiferromagnetic phase. Neutron and resonant x-ray
scattering results of Keimer et al.\cite{keimer1} have been
interpreted as evidence for orbital fluctuations, consistent with
orbital liquid model of Khaliullin and Maekawa \cite{khaliullin1}.
It has been pointed out \cite{keimer1} that earlier Raman results
of Reedyk at al.\cite{reedyk1}, where large background and Fano
shape of the phonon line near 300~cm$^{-1}$ have been observed,
may also indicate orbital fluctuations coupled to lattice
vibrations. Some recent experimental results on the temperature
dependence near T$_N$ of neutron and x-ray diffraction, heat
capacity and infrared spectra \cite{cwik1,hemberger1}, however,
provide evidence for noticeable deformation of TiO$_6$ octahedra
and structural anomaly near the antiferromagnetic ordering, which
indirectly supports the concept of orbital ordering.

The observation in the Raman spectrum of {\it insulating} rare
earth titanates of structureless background and Fano interference
is highly unusual. At the same time, it is well known that
transport and magnetic properties of LaTiO$_3$
\cite{okada1,taguchi1,meijer1} depend crucially on sample's
stoichiometry. Critical dependence on stoichiometry may be
expected also for the Raman spectra. Indeed, anomalous variation
of phonon Raman intensities and linewidths has been reported at
50~K for LaTiO$_{3+\delta/2}$ near the metal-to-Mott-insulator
transition at $0.01 < \delta < 0.04$  \cite{katsufuji1}. The room
temperature dc resistivity ($\rho = 0.02$~$\Omega$cm) of the
LaTiO$_3$ sample used in the Raman experiments of Reedyk et
al.\cite{reedyk1} is much lower than that reported for nearly
stoichiometric samples ($\rho > 0.5$~$\Omega$cm
\cite{okada1,taguchi1}) and rather corresponds to $\delta = 0.04$.
Therefore, it is of definite interest to examine the Raman spectra
of stoichiometric LaTiO$_3$ in a broad temperature range including
N\'{e}el temperature T$_N$. It is plausible to expect that the
variation of the Raman spectra with decreasing temperature below
T$_N$ will provide additional information on the controversial
issues of Fano interference and magnetic-order-induced orbital
ordering.

In this paper we present polarized temperature-dependent Raman
spectra of stoichiometric LaTiO$_3$ (T$_N = 150$~K) between 6 and
300~K. At room temperature, in contrast to Ref.\cite{reedyk1},
neither strong background scattering nor Fano shape of the Raman
lines is observed. The temperature shift of some Raman lines
exhibits clear anomaly below T$_N$: a signature for structural
rearrangement. We discuss the assignment of the Raman lines to
definite phonon modes and the nature of structural changes. The
broad Raman band, which appears in the antiferromagnetic phase, is
assigned to two-magnon scattering.

LaTiO$_3$ samples were prepared using La$_2$O$_3$ (99.99\%),
TiO$_2$ (99.99\%) and Ti$_2$O$_3$ (99.99\%) as starting materials.
La$_2$O$_3$ was heat treated at 1300$^\circ$C for 24h and TiO$_2$
was dried for 24h at 120$^\circ$C before use. Stoichiometric
amounts of La$_2$O$_3$, Ti$_2$O$_3$ and TiO$_2$ were mixed and arc
melted under argon to form black bulk LaTiO$_3$. X-ray diffraction
pattern at room temperature revealed orthorhombic structure with
lattice parameters $a=5.61$~\AA,  $b=7.91$~\AA, and $c=5.63$~\AA,
in agreement with earlier reports \cite{cwik1,fritsch1}. It is
known that the magnetic transition temperature, T$_N$, is very
sensitive to the oxygen content \cite{taguchi1} and it rapidly
shifts to lower T if the oxygen composition exceeds the
stoichiometric value of 3 \cite{meijer1}. Therefore, the value of
T$_N$ is a precise measure of the oxygen stoichiometry in
LaTiO$_3$. The weak ferromagnetism is due to the asymmetric
Dzyaloshinsky-Moriya exchange interaction and can easily be picked
up in dc-susceptibility measurements.

\begin{figure}
\includegraphics[width=3in]{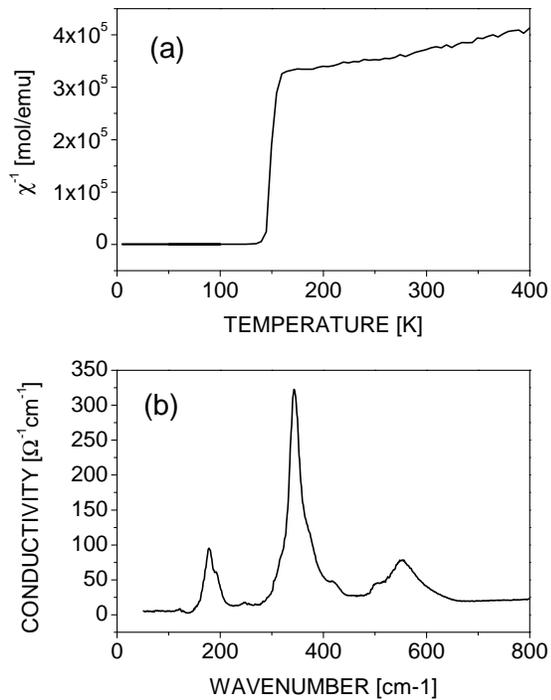}
\caption{(a) Inverse magnetic susceptibility of LaTiO$_3$ measured
 at 50 Oe as a function of temperature; (b) Room temperature conductivity,
 obtained from the near-normal reflectance data.}
\end{figure}

For our sample the SQUID magnetometry was employed to measure the
magnetic transition temperature. Fig. 1(a) shows the inverse
susceptibility data measured at 50 Oe in the temperature range
between 5 and 400~K. A sharp drop occurs at 150~K, in excellent
agreement with the best available data for stoichiometric single
crystals \cite{fritsch1}.
Another supportive evidence, which points to almost perfect sample
stoichiometry, is the room temperature conductivity, obtained by
Kramers-Kronig transformation of near-normal reflectance. It
extrapolates [Fig. 2(b)] to the dc-value of
$4.5\pm0.5~\Omega^{-1}$cm$^{-1}$ ($\rho=0.22 \pm 0.03~\Omega$cm),
which corresponds to $\delta<$~0.01 \cite{taguchi1}.

Raman spectra were collected under microscope (focus spot size
1-3~$\mu$m, $\lambda_{exc} = 514.5$~nm or 632.8~nm) from freshly
cleaved or as-grown surfaces of the bulk material. The
crystallographic orientation of the surface was not known but in
most cases the spectra taken with parallel (HH) and crossed (HV)
polarizations of incident and scattered radiation were totally
polarized: an indication that the surface coincides with one of
the main crystallographic planes ($ab$, $bc$ or $ac$). For
temperature-dependent measurements the sample was mounted in a
liquid helium cryostat. As the Raman signals were extremely low,
re\-la\-ti\-vely high incident laser power ($\approx 5$~mW) was
used, which resulted in some heating of the microprobe spot.

Fig.~2 shows the polarized Raman spectra of LaTiO$_3$ as obtained
at room temperature from five different spots. Three Raman lines
are clearly pronounced in parallel (HH) scattering configuration
at 133, 252 and 296~cm$^{-1}$ and four other lines are seen in
crossed (HV) configuration at 182, 418, 465, and 655~cm$^{-1}$.
The phonon line positions are close to those reported by Reedyk et
al. (see Fig.~2 in \cite{reedyk1}), but the background scattering
is much weaker and the lines are narrower. The most significant
difference is the observation of two clearly distinguishable
symmetric lines at 252 and 296~cm$^{-1}$ instead of one broader
asymmetric band between 220 and 300~cm$^{-1}$. Taking into account
that for LaTiO$_{3+\delta/2}$ one expects an increase of the
electronic background, phonon line intensity and phonon line width
with increasing $\delta$ \cite{katsufuji1}, the puzzling "Fano
shaped" band reported by Reedyk et al.\cite{reedyk1} seems to have
simpler explanation as a complex band, consisting of two
relatively broad symmetric lines centered at 252 and
296~cm$^{-1}$.

\begin{figure}
\includegraphics[width=3in]{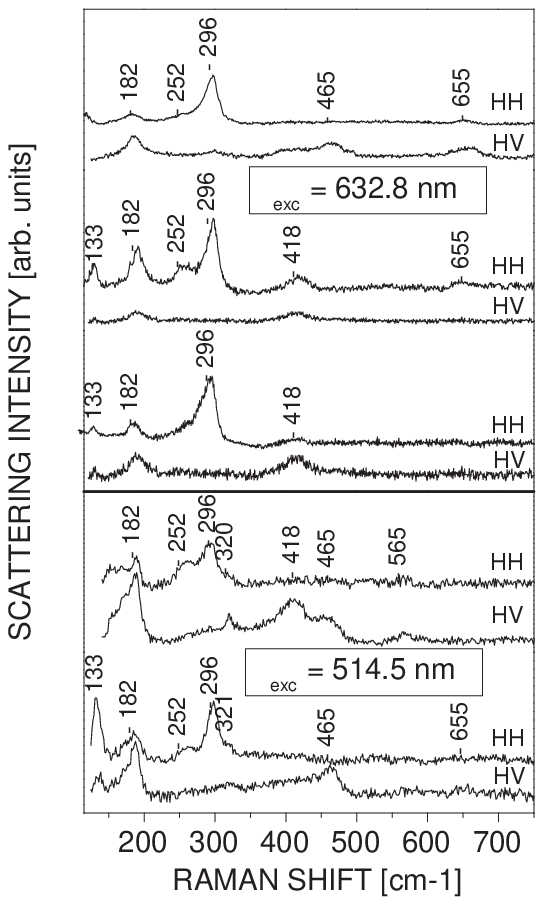}
\caption{Raman spectra of LaTiO$_3$ obtained at room temperature
with parallel (HH) and crossed (HV) scattering configurations from
freshly cleaved (a,c,d) and as grown (b) surfaces.}
\end{figure}

\begin{figure}
\includegraphics[width=2.6in]{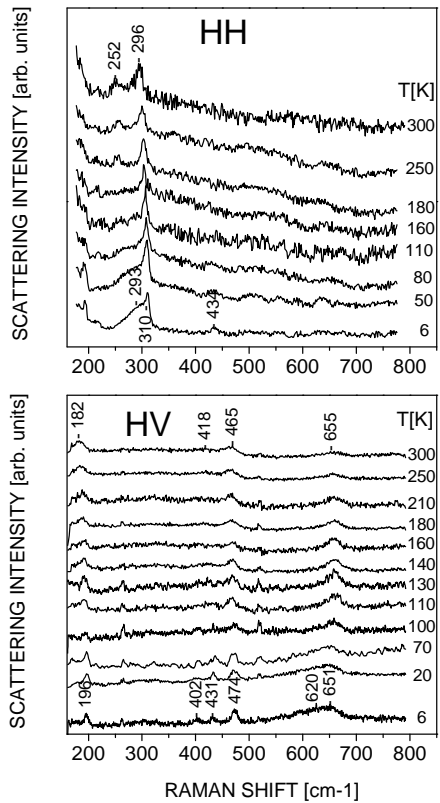}
\caption{Variations with temperature of the HH and HV Raman
spectra of LaTiO$_3$. Due to local laser heating the actual
temperature is higher than the nominal one.}
\end{figure}

The variations with temperature of the HH and HV Raman spectra are
shown in Fig.3.  Upon lowering temperature some of the lines
[182~cm$^{-1}$(HV), 296~cm$^{-1}$(HH), 465~cm$^{-1}$(HV)] exhibit
normal monotonous narrowing and hardening to 196, 310 and
474~cm$^{-1}$, respectively. The line at 252~cm$^{-1}$ decreases
in intensity and cannot clearly be detected as the nominal
temperature approaches T$_N$. Instead, in the AFM phase a
relatively broad line arises between 250 and 300~cm$^{-1}$. At low
temperatures two additional lines are clearly pronounced at 402
and 431~cm$^{-1}$ in the HV spectra. The line, which exhibits
anomalous temperature behavior, is the one at 655~cm$^{-1}$. As
illustrated in Fig.~4, with lowering temperature between 300~K and
130~K this mode hardens and increases in intensity. Upon further
cooling, however, it moves back to lower wavenumbers and merges
with an arising new broad band centered at about 620~cm$^{-1}$.
The position of this latter band is independent of temperature
within the experimental error ($\pm$ 5~cm$^{-1}$) and it's
intensity increases much faster compared to intensity decrease of
the 655 cm$^{-1}$ phonon line.

In order to assign the observed Raman lines to definite phonon
modes we performed lattice dynamical calculations (LDC) using a
shell model, which has been applied earlier for isostructural
YMnO$_3$ and LaMnO$_3$ \cite{iliev1}. To evaluate the effect of
structural changes, identical calculations were done using the
neutron diffraction data of Cwik et al. \cite{cwik1} for atomic
positions at 8~K, 155~K an 293~K. The LDC results showed that both
the predicted frequencies and shapes of the phonon modes of
LaTiO$_3$ and LaMnO$_3$ are very close. The three HH lines can
unambiguously be assigned to $A_g$ modes involving mainly motions
of La along $z$ (133~cm$^{-1}$), in-phase rotations around $y$ of
neighboring (along $y$) TiO$_6$ octahedra (252~cm$^{-1}$), and O1
motions in the $xz$ plane (295~cm$^{-1}$), respectively. The
assignment of HV lines is less straightforward as to each
experimentally observed line one can juxtapose $B_{1g}$, $B_{2g}$
or $B_{3g}$ mode of close predicted frequency. Whatever is the
choice, the line at 182~cm$^{-1}$ corresponds to a mode involving
mainly motions of La and the line at 655~cm$^{-1}$ - to in-phase
($B_{2g}$) or out-of-phase ($B_{1g}, B_{3g}$) stretchings of
TiO$_6$ octahedra.

\begin{figure}
\includegraphics[width=2.5in]{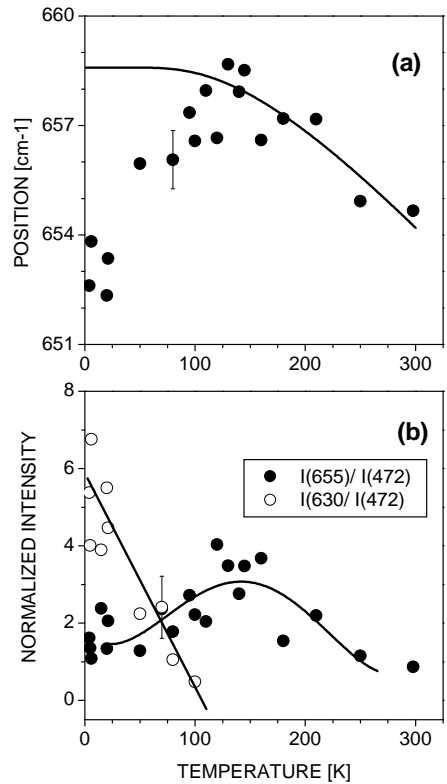}
\caption{Temperature dependence of the phonon line position (a)
and normalized intensities (b) of the two high frequency bands.
Solid line in (a) shows the behavior expected for a standard
anharmonic phonon decay. Lines in (b) are guide to the eye.}
\end{figure}

The comparison of frequencies calculated using structural data for
at 8~K, 155~K, and 293~K provides evidence that the softening of
the 655~cm$^{-1}$ mode below T$_N$ is related to the structural
changes induced by magnetic ordering \cite{cwik1}. These changes
include elongation of TiO$_6$ octahedra along $c$ (in $Pnma$
notations) by 0.2\% and shrinking along $a$ by 0.3\% and an
anomalous shortening of the Ti-O1 distance near T$_N$. Indeed, the
LDC results predict hardening of all Raman modes except for the
$B_{1g}$, $B_{2g}$ and $B_{3g}$ modes near 650~cm$^{-1}$. The
latter modes have maximum wavenumber at 155~K and then either
soften by $\sim 1$~cm$^{-1}$ ($B_{1g}, B_{3g}$), or remain
unchanged ($B_{2g}$) at 8~K. Interestingly, similar softening
below T$_N$ has also been observed for the corresponding $B_{2g}$
mode in isostructural LaMnO$_3$ \cite{podobedov1,iliev2}. Another
result that follows from the comparison of LCD data of LaTiO$_3$
at different temperatures is that the HH line at 252~cm$^{-1}$ at
300~K and the broad band at $\approx 290$ ~cm$^{-1}$ at 6~K most
likely correspond to the same "soft" mode. It is worth noting that
all predicted temperature shifts are by factor 3 smaller than the
experimentally observed ones. The reason for this discrepancy is
not clear.

\begin{figure}
\includegraphics[width=3in]{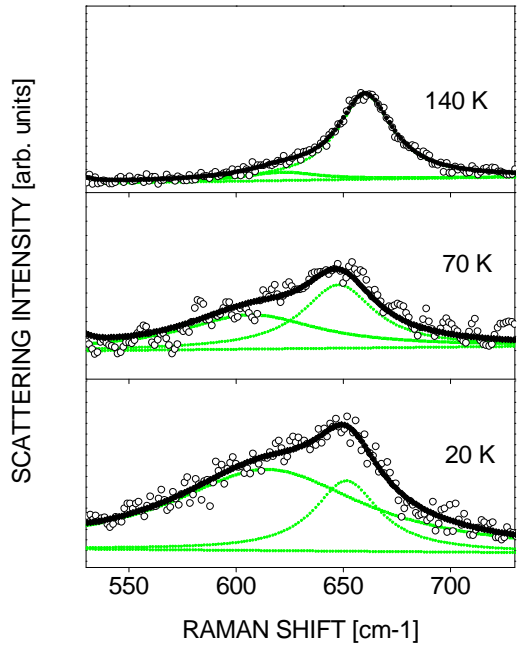}
\caption{Evolution of the two-magnon band upon lowering
temperature. The experimental data are represented by open
points.}
\end{figure}

The broad structure between 570 and 650~cm$^{-1}$ appears only
below T$_N$ and strongly increases with lowering temperature (see
Fig.~5) thus identifying itself as related to magnetic
excitations. The magnon dispersion curve for LaTiO$_3$ along the
pseudocubic [111] direction was measured by Keimer et
al.\cite{keimer1} and fitted by the expression  $\hbar\omega =
zSJ\sqrt{1.005-\gamma^2}$, where $\hbar\omega$ is the magnon
energy, $z=6$ is the coordination number, $S=1/2$ is the Ti spin,
$J=15.5\pm1.0$~meV  is the nearest-neighbor superexchange energy,
$\gamma = \frac{1}{3}[\cos(q_xa)+\cos(q_ya)+\cos(q_za)]$. The
zone-center magnons ($\gamma =1,\ \hbar\omega = 3.3$~meV) are far
below our range of measurement and only second-order magnetic
scattering is expected. The intensity of the two-magnon scattering
is determined by the magnitude of the two-magnon density of
states, which has maximum at the zone boundary, and by the
interaction of the two magnons created in the scattering process.
The latter interaction results in creation of a "bound state",
which decreases the two-magnon energy by $J$ compared to the sum
of individual magnon energies at the boundary $2zSJ$
\cite{elliot1,hayes1}. Therefore, the maximum of two-magnon
scattering for LaTiO$_3$ is expected at $\hbar\omega_{2M} = 2zSJ -
J =  5J$. The position of the broad line in the spectra is $620
\pm 20$~cm$^{-1}$ ($76.9\pm2.5$~meV) yields $J = 15.4 \pm
0.5$~meV, in excellent agreement with the results of neutron
scattering experiments \cite{keimer1}.

\acknowledgments This work is supported in part by the state of
Texas through the Texas Center for Superconductivity and Advanced
Materials.

\end{document}